\documentclass[
aps,
prb,
twocolumn,
showpacs,
superscriptaddress,
groupedaddress,
nofootinbib,
longbibliography,
floatfix,
]{revtex4-2}  %

\usepackage[pdftex]{graphicx}
\usepackage[english]{babel}
\usepackage{amsmath}
\usepackage{soul}
\usepackage{bbold}
\usepackage{txfonts}
\usepackage{hyperref}
\usepackage{color}

\begin{document}

\title{Higher-dimensional Euclidean and non-Euclidean structures in planar circuit
quantum electrodynamics }
\author{Alberto Saa}
\email{asaa@ime.unicamp.br}
\affiliation{
Department of Applied Mathematics,
University of Campinas (UNICAMP),  13083-859 Campinas, SP, Brazil.}
\author{Eduardo Miranda}
\email{emiranda@ifi.unicamp.br}
\affiliation{Institute of Physics Gleb Wataghin, University of Campinas (UNICAMP),   13083-859 Campinas, SP, Brazil}
\author{Francisco Rouxinol}
\email{rouxinol@ifi.unicamp.br}
\affiliation{Institute of Physics Gleb Wataghin, University of Campinas (UNICAMP),   13083-859 Campinas, SP, Brazil}

\date{\today}

\begin{abstract}
We demonstrate that a recent proposal for simulating planar hyperbolic lattices using circuit quantum electrodynamics can be extended to include higher-dimensional lattices in both Euclidean and non-Euclidean spaces by allowing circuits that involve more than three polygons at each vertex. The quantum dynamics of these circuits, which we are developing with current technology, are governed by effective tight-binding Hamiltonians that correspond to higher-dimensional Kagom\'e-like structures (such as $n$-dimensional zeolites). These structures are known for exhibiting strong frustration and flat bands. We analyze the spectra of both hyperbolic and positive-curvature lattices and derive exact expressions for the fraction of flat-band states. Our findings significantly broaden the possibilities for realizing non-Euclidean geometries using circuit quantum electrodynamics, a research direction we are actively pursuing in microwave-guide circuits constructed with sputtered niobium films on silicon substrates.
 \end{abstract}

\maketitle

\section{Introduction} 
There has been a long history of cross-pollination between geometry and various areas of physics \cite{atiyah_geometry_2010} .
Geometry is at the base of general relativity  and cosmology, leading also to surprising semiclassical  effects such as Hawking radiation. The difficulty of directly observing these subtle quantum effects in a gravitational context has spurred the search for analogues in condensed matter systems \cite{steinhauer_observation_2016,eckel_rapidly_2018,kolobov_observation_2021,philbin_fiber_optical_2008,weinfurtner_measurement_2011}.
Non-flat geometries, however, have proved fruitful even in situations that are not gravity-related.
A prime example is geometric frustration. The optimal {\em local} packing of hard spheres in an icosahedral structure cannot be periodically extended in Euclidean space. It is, however, compatible with periodicity in hyperbolic space, which can then serve as a starting point. The real system can then be approximated and analyzed by introducing defects into the pristine hyperbolic idealization (see, e.g., \cite{tarjus_frustration_based_2005} for a review).
Other examples of this cross-fertilization include the control of infrared singularities in classical and quantum field theories in hyperbolic space \cite{callan_infrared_1990},
the anti-de Sitter/conformal field theory duality \cite{maldacena_large_n_1999}, 
phase transitions in curved spaces \cite{krcmar_tricritical_2008,mnasri_critical_2015,breuckmann_critical_2020}, and hyperbolic surface codes for quantum computation \cite{breuckmann_constructions_2016}, 
among many others.

More recently, the flexibility of the design of circuit quantum electrodynamics (cQED) \cite{koch_time-reversal-symmetry_2010,nunnenkamp_synthetic_2011,Blais2020} has enabled the experimental realization of hyperbolic lattices using planar microwave circuits \cite{kollar_hyperbolic_2019,kollar_line-graph_2020, boettcher_quantum_2020}. In these systems, multiple microwave resonators are capacitively coupled to form an artificial photonic lattice. The photon dynamics can be effectively described by a tight-binding model in a hyperbolic plane.
{This important achievement has stimulated some recent advances such as the formulation of a band theory in hyperbolic lattices \cite{maciejko_hyperbolic_2021} or proposals for the realization of topological phases \cite{yu_topological_2020}.}

{A promising direction that follows from these results is the integration of superconducting qubits into these architectures, allowing for the implementation of fully interacting models~\cite{houck_-chip_2012,zhuetal2021}. As far as we know, however, such experiments have not yet been reported for either the $q=3$ or $q>3$ cases. 
Although there are reports describing measurements involving $N=3$ sites with qubits in a Ph.D. thesis~\cite{thesis}, no peer-reviewed publications have resulted from those results. 
Other recent works~\cite{scigliuzzo_extensible_2022,zhang_superconducting_2023,obrien_circuit-qed_2025} explore lattices with superconducting qubits, but in distinct network topologies.
} 
 
\begin{figure}[t]
\includegraphics[width=1\linewidth]{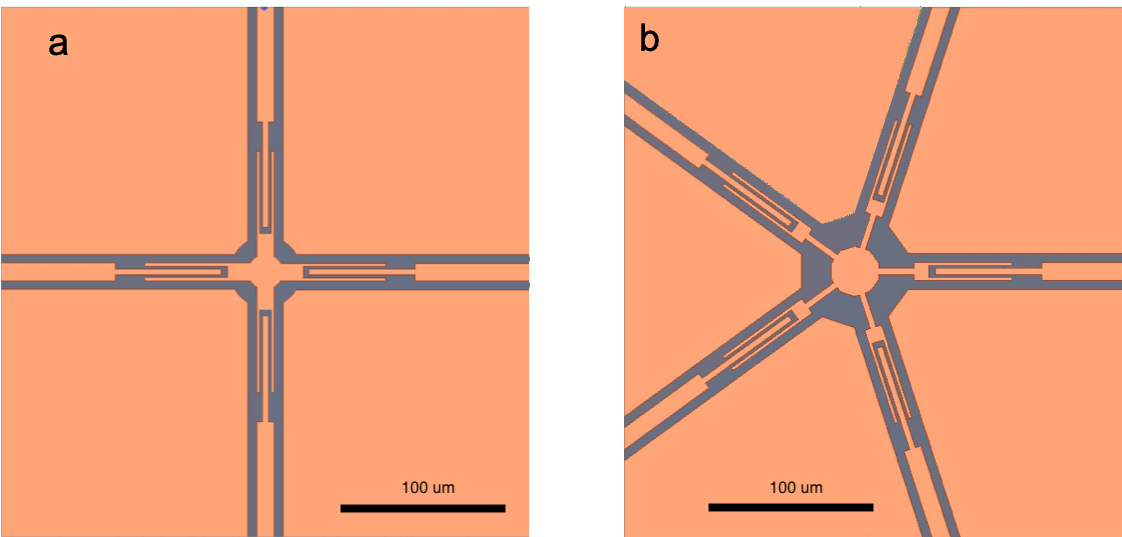} 
\caption{\label{fig.circuit} Proposed planar \(q\)-leg capacitive devices coupling the resonators for (a) \(q=4\) and (b) \(q=5\).
We are building these circuits  using standard micro-fabrication techniques. In the 4-leg capacitor (a), for instance, the capacitance between any pair of legs is 374pF, with deviations smaller than 0.01pF.  The generic case with $q$ symmetrical legs follows analogously as a  star-shaped configuration with $q$ leaves. See the Appendix \ref{apa}
further construction details. }
 \end{figure}

A severe limitation of the systems built so far is their confinement to strictly two-dimensional lattices. Indeed, the planar layout of the circuits seems, at first, to preclude a higher-dimensional setup. We propose in this paper a way to overcome this limitation by increasing the connectivity of the microwave resonators. This is achieved by means of a capacitive coupling design that can symmetrically couple $q > 3$ resonators with equal strength, a $q$-leg capacitor that can be easily constructed with present technology (see Fig.~\ref{fig.circuit}). 
As a result, even though the device layout is contained within the usual planar design, the effective dimension of the underlying dynamics is greater than 2, forming a so-called 
$n$-zeolite framework \cite{bonchev2000chemical}.
This enlarges considerably the range of possible applications and opens the possibility of exploring different hyperbolic structures with flat bands, as we will show. 
It also affords the flexibility of generating a spatially varying connectivity and, consequently, a non-homogeneous geometric configuration.
{ Besides exploring this new design in both cases of positive and negative curvature, we also derive some results regarding the spectra of these systems, such as a generic expression for the fraction of flat-band states
and some bounds for the largest eigenvalues 
of full and half-wave modes.}
\begin{figure}[t]
\includegraphics[width=0.49\linewidth]{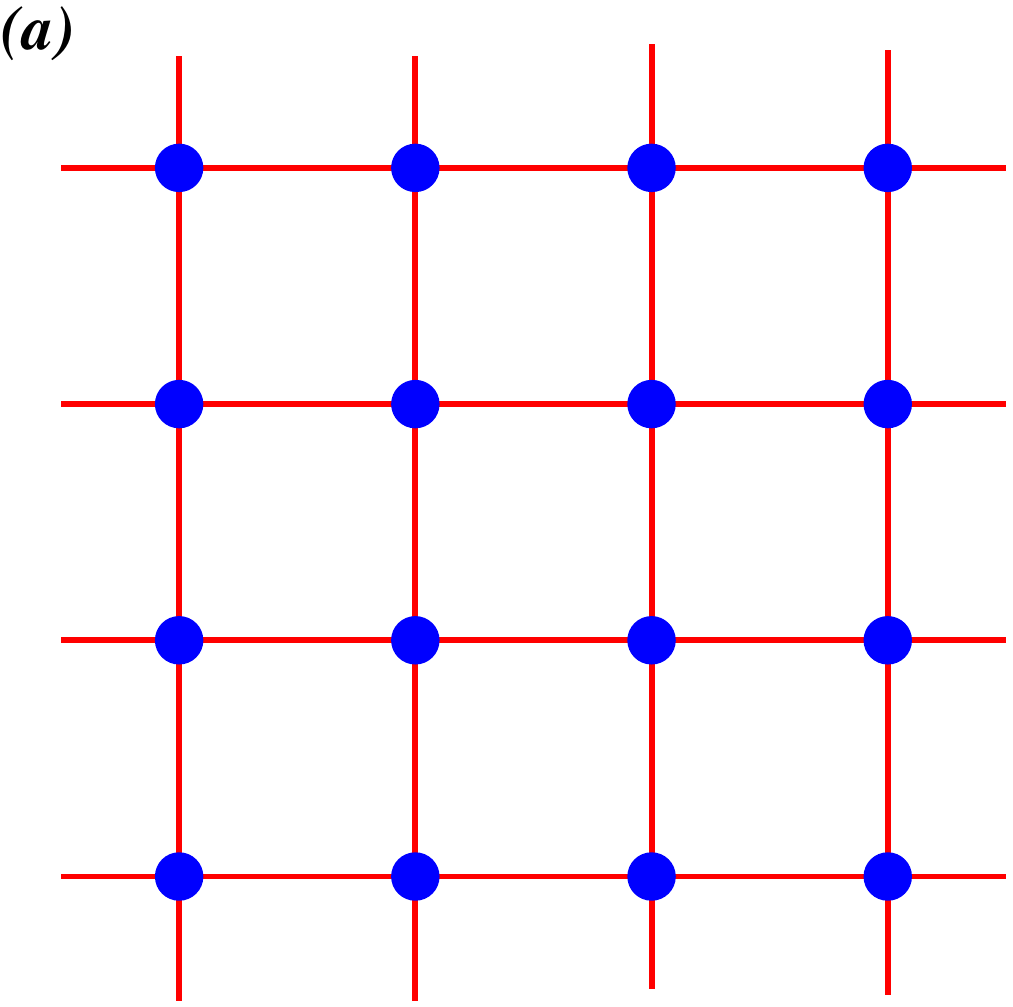}
\includegraphics[width=0.49\linewidth]{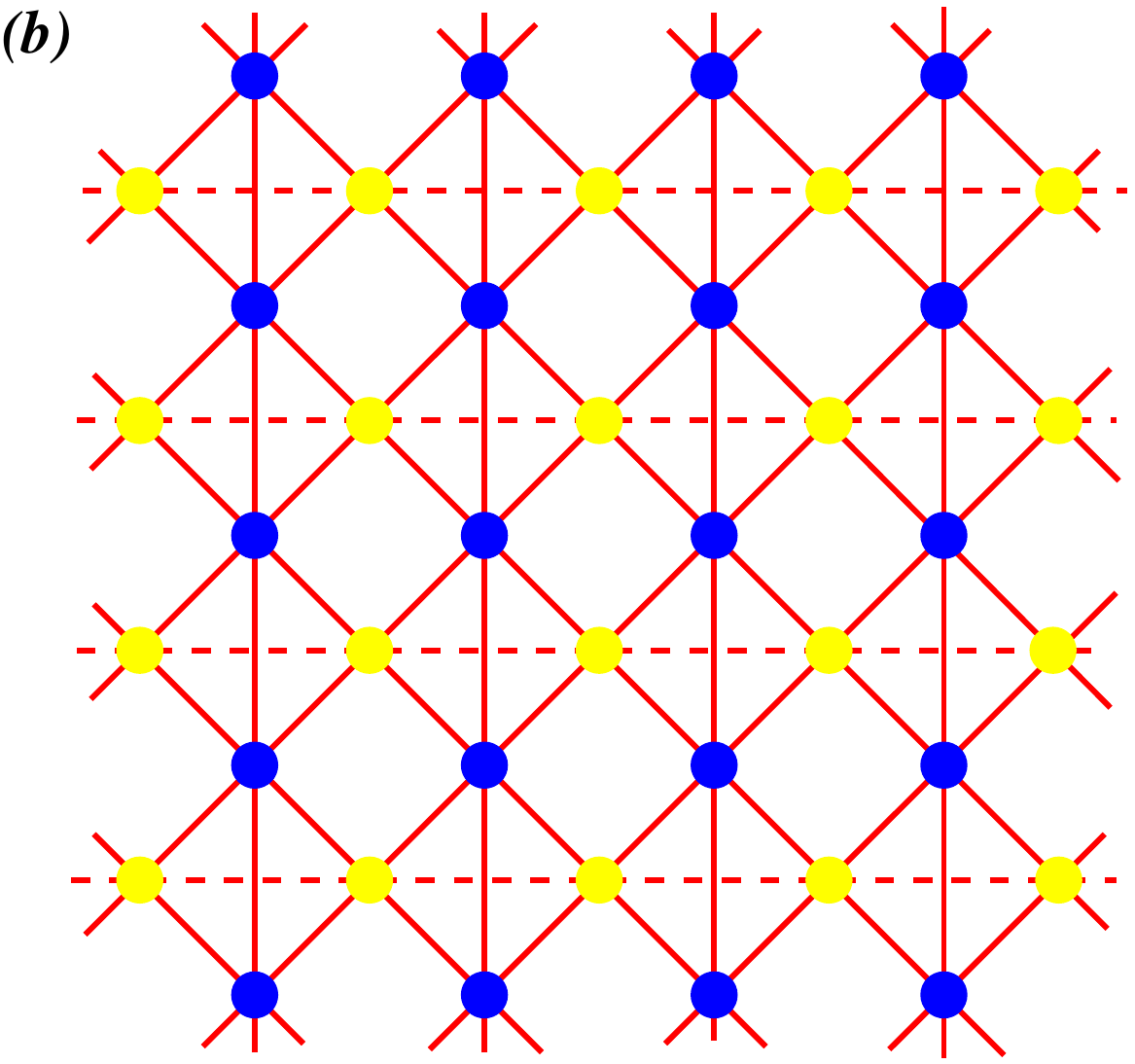}\\
\vspace{0.5cm}
\includegraphics[width=0.49\linewidth]{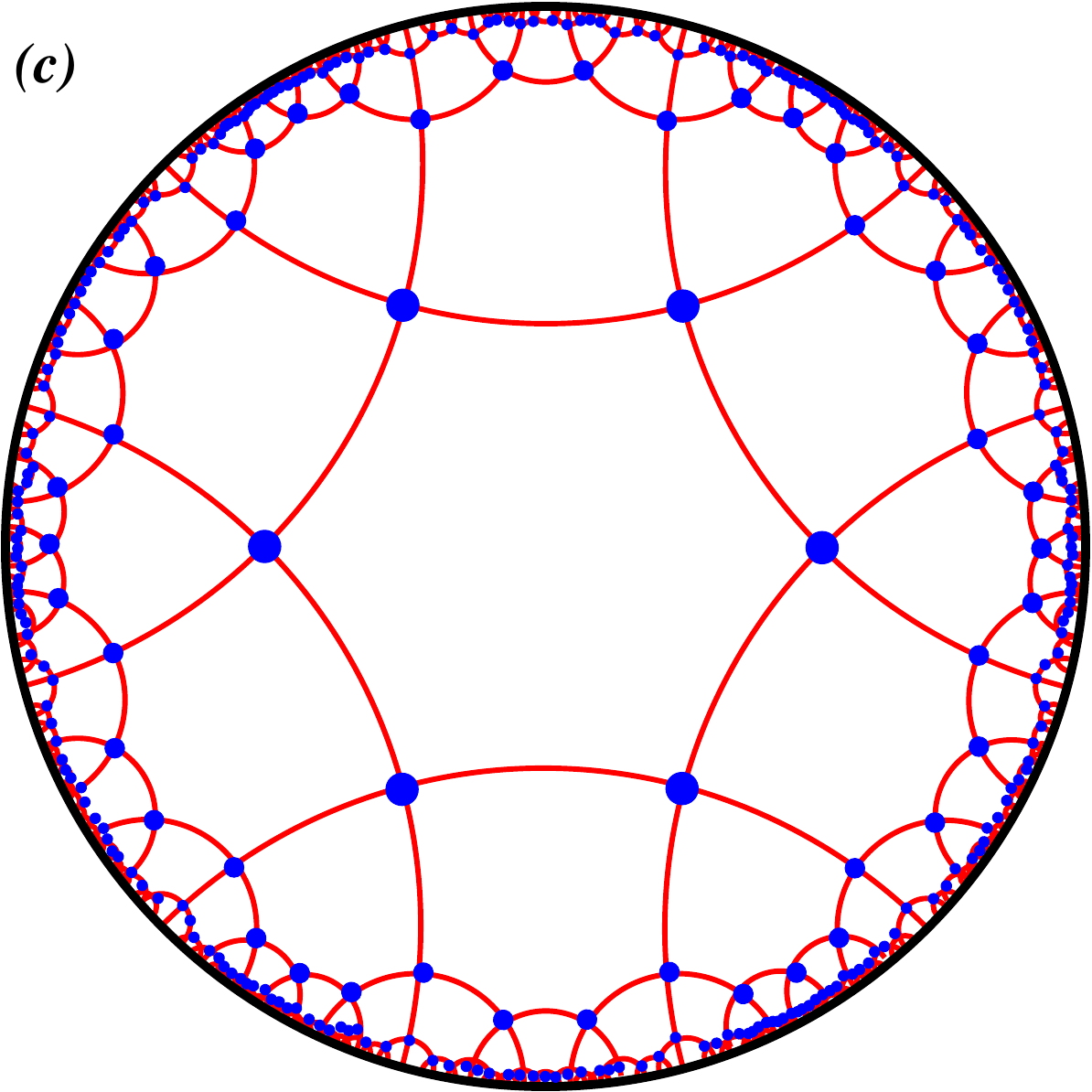}
\includegraphics[width=0.49\linewidth]{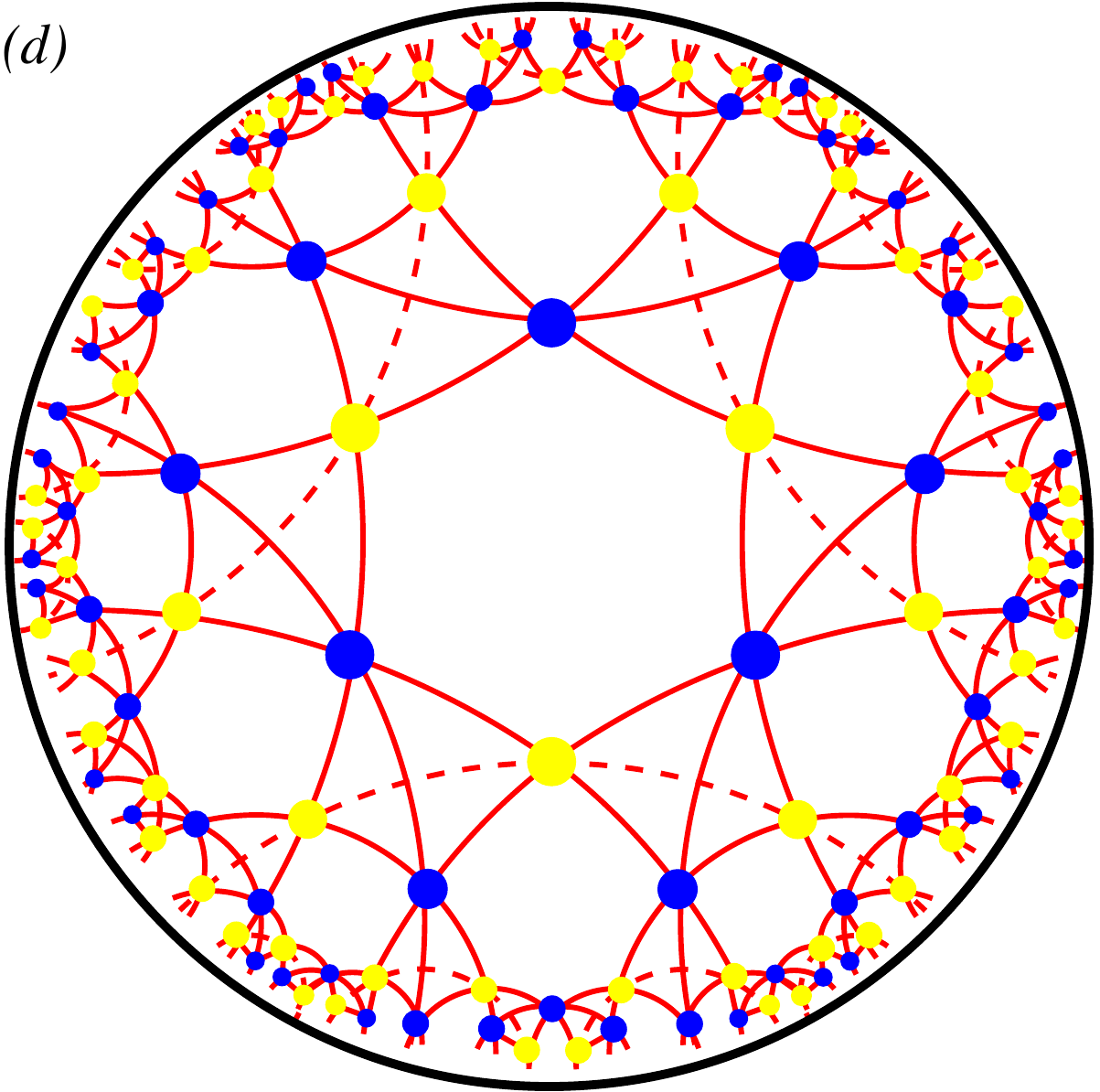}
\caption{\label{fig1} {Some examples of tilings with  $q=4$ and their associated line graphs. (a) The usual square $\{4,4\}$ tiling  of $\mathbb{E}^2$ and (b) its associated line graph, which is equivalent to a single layer of corner-sharing  tetrahedra 
($3$-zeolites) in $\mathbb{E}^3$, with the blue and yellow vertices located in two parallel planes and seen from a perpendicular viewpoint. (c) The hexagonal  $\{6,4\}$ tiling  of $\mathbb{H}^2$ and (d) its associated line graph, which can be viewed in an analogous way:  a single layer of corner-sharing tetrahedra in $\mathbb{H}^3$ (or in $\mathbb{H}^2\times \mathbb{R}$), viewed from above. Note that a  layered geometrical realization of the line graphs of (b) and (d) are only available for $\{p,4\}$ tilings with  even $p$,
since the disposition of the blue and yellow vertices in two parallel planes is only possible if the line graph
is bipartite. 
 We employ here Schl\"afli's $\{p,q\}$-notation for two-dimensional regular tilings (see text).} } 
\end{figure}
{In the following section, we will provide a brief review of some fundamental results in circuit quantum electrodynamics (cQED), including their layouts and the underlying lattices. Section III will focus on higher-dimensional geometries and the circuits associated with them. Our main results will be presented in Section IV, while the Final Remarks will be discussed in Section V.  Some construction details for the  $q$-leg capacitor are included in Appendix A.}

\section{Superconducting Circuits and Lattices}

Let us briefly review some of the basics of cQED \cite{kollar_hyperbolic_2019, kollar_line-graph_2020, boettcher_quantum_2020}. These photonic systems consist of identical quantum microwave resonators disposed along the edges of a layout lattice. Each vertex of the lattice is a $q$-leg capacitor, responsible for the symmetric pairwise coupling between the $q$ resonators meeting at that vertex. This defines a lattice called the layout graph $G$ 
[see Figs.~\ref{fig1}(a) and \ref{fig1}(c)]. The underlying quantum dynamics of the system is governed by a tight-binding Hamiltonian 
\begin{equation}
\label{ham}
H = H_0 + H_I =  \omega_0 \sum_ia^\dagger_ia_i - \sum_{\langle i,j\rangle}t_{ij} \left( a^\dagger_i  a_j +  a^\dagger_ja_i \right),
\end{equation}
where $\omega_0$ is the resonator frequency. The off-diagonal term $H_I$ describes the hopping (with amplitude $t_{ij}$) of photons between resonators induced by the capacitors. It is clear that the sites of the Hamiltonian of Eq.~\eqref{ham} should be taken to be the midpoints of the edges of the layout lattice, and its connectivity is determined by the capacitors. This underlying lattice is called the line graph, which we will denote by $L(G)$
[see Figs.~\ref{fig1}(b) and \ref{fig1}(d)]. 

In order to describe either type of graph we will use Schl\"afli's $\{p,q\}$-notation for two-dimensional regular tilings. It denotes a tiling with $p$-regular polygons, or $p$-gons, disposed so that $q$ of them meet at every corner.  A regular hyperbolic tiling  requires only
\begin{equation}
 \label{tau}
 \tau = (p-2)(q-2)>4 ,
\end{equation}
with no further restrictions on the polygons besides being convex and regular. Hence, there are (countably) infinitely many regular tilings of the hyperbolic plane $\mathbb{H}^2$, in contrast to the possible tilings of the usual Euclidean plane $\mathbb{E}^2$ and the sphere $\mathbb{S}^2$, for which $\tau=4$ and $\tau<4$, respectively. The particular choice of $q=3$ for the layouts explored in \cite{kollar_hyperbolic_2019} leads to line graphs that are Kagom\'{e} lattices of corner-sharing triangles. These are highly frustrated two-dimensional lattices with characteristic flat bands in their tight-binding spectra \cite{kollar_hyperbolic_2019,kollar_line-graph_2020,boettcher_quantum_2020}.

The absolute value of $t_{ij}$ is homogeneous in the lattice, but its sign may vary. Two sets of modes arise naturally in this system, which should be treated separately \cite{kollar_line-graph_2020}. The first are the so-called full-wave or symmetrical modes, for which the sign of $t_{ij}$ is the same for all resonator pairs $(i,j)$. In this case, we can write, in matrix notation,
\begin{equation}
\label{H_s}
	H_I = H_s =-t A_{LG},
\end{equation}
where $A_{LG}$ stands for the adjacency matrix of the corresponding line graph. The second set of modes are the half-wave or antisymmetrical modes, for which the sign of  $t_{ij}$ varies throughout the lattice. The signs of $t_{ij}$ depend on a chosen orientation of the edges of the layout graph $G$. This means that each edge of $G$ should be assigned a head vertex and a foot vertex. We can then write
\begin{equation}
\label{H_a}
H_I = H_a =   -t A_{LG}^*,
\end{equation}
where the matrix $A_{LG}^*$ is such that its entries are \cite{kollar_line-graph_2020}
\begin{equation}
 \left[ A_{LG}^*\right]_{ij} = \left\{
\begin{array}{rc}
1, & {\rm if\ }  e_i^+ = e_j^+ \ \mathrm{or} \ e_i^- = e_j^- , \\
-1, &  {\rm if\ } e_i^+ = e_j^- \ \mathrm{or} \ e_i^- = e_j^+ , \\
0, & {\rm otherwise,}
\end{array}
 \right. \label{orient}
\end{equation}
where $e_i^\pm$ denotes the head $(+)$ and foot $(-)$ of the oriented edge whose midpoint is $i \in L(G)$ and the comparisons in Eq.~\eqref{orient} refer to the vertex shared by the edges $i$ and $j$. The matrix $A_{LG}^*$ is the adjacency matrix of the so-called signed line graph of the layout (see, e.g., \cite{zaslavsky_signed} for further details on signed graphs). Although its entries depend on the chosen orientation for $G$, a change of orientation of any edge of $G$ (a so-called switching operation) preserves the spectra of Eq.\eqref{H_a}. {Actually, a switching operation corresponds to a gauge transformation of the Hamiltonian~\eqref{ham}, which obviously preserves the spectra. In order to see this, let us
consider an edge of the layout lattice. Its midpoint is a lattice site of the line graph, which we will identify as site 0. Now perform a switching operation, {\em i.e.}, change the orientation of this edge while preserving all the other edge orientations. From Eq. (\ref{orient}), it is easy to see that this reverses the signs of the hopping amplitudes $t_{0j}$ between site 0 and its nearest neighbors in the line graph, while leaving all the other hopping amplitudes unchanged. In other words, the only change in the tight-binding Hamiltonian is in the terms
\begin{equation}
    \sum_{\langle0j\rangle} t_{0j} \left(a^{\dagger}_0 a^{\phantom{\dagger}}_j+a^{\dagger}_j a^{\phantom{\dagger}}_0\right)
    \to
    -\sum_{\langle0j\rangle} t_{0j} \left(a^{\dagger}_0 a^{\phantom{\dagger}}_j+a^{\dagger}_j a^{\phantom{\dagger}}_0\right).
\end{equation}
Now, this is a particular case of a lattice gauge transformation (a more general $U(1)$ transformation would involve distinct phases $e^{i\phi_{ij}}$), which can be easily undone through the   canonical transformation of the creation and annihilation operators given by
$
    a^{\phantom{\dagger}}_0\to -a^{\phantom{\dagger}}_0, \ \ a^{\dagger}_0\to -a^{\dagger}_0.
$}

\section{Higher dimensional geometries.}

Circuits based on
$\{p,q\}$ tilings with $q>3$, naturally lead to some effective higher dimensional structures. Fig.~\ref{fig1} depicts, for example, the $\{4,4\}$ and $\{6,4\}$ tilings of $\mathbb{E}^2$ [(a)] and $\mathbb{H}^2$ [(c)], and their associated line graphs [(b) and (d)], respectively. Note the higher-dimensional ``cages'' (tetrahedra) of the line graphs. Our proposal for the construction of these lattices depends critically on the existence of efficient implementations of symmetric planar capacitors with $q$-legs. Fig.~\ref{fig.circuit} depicts a possible star-shaped construction for these devices based on the usual techniques of cQED, see  Appendix \ref{apa} for further details. In such a device, \emph{any} pair of legs experiences the same mutual capacitance, not only adjacent ones.

In general, the quantum dynamics of a $\{p,q\}$-layout circuit will be determined by its line graph (see Fig.~\ref{fig1}): the full and half-wave modes will be governed, respectively, by Eqs.~\eqref{H_s} and \eqref{H_a}. Such line graphs are composed of vertex-sharing subgraphs, each of which is a regular $(q-1)$-simplex. A regular $n$-simplex is the convex hull (polyhedron) of $n+1$ equidistant points in some $n$-dimensional space. For the $q=4$ case of Fig.~\ref{fig1} this simplex is a regular tetrahedron.
Note that the simplices are regular due to the symmetry of the capacitive coupling and the homogeneity of the resonators. In general, the line graph associated with a $\{p,q\}$-layout with symmetric couplings will be a  regular graph with $2(q-1)$ edges per vertex, corresponding to a structure of corner-sharing identical $(q-1)$-simplices. Such structures of corner-sharing identical $n$-simplices are known in the literature as $n$-dimensional zeolites \cite{bonchev2000chemical}. Besides, its geometrical realization as an embedding, \emph{if possible}, clearly demands at least a $(q-1)$-dimensional background space, which cannot be Euclidean unless the original layout is also Euclidean. Again, for $q=4$, we need 3 dimensions, as seen in Fig.~\ref{fig1}(d).

It should be emphasized, however, that not all corner-sharing $(q-1)$-simplex frameworks corresponding to a $\{p,q\}$-tiling line graph will admit layered embeddings as those depicted in Fig. \ref{fig1}. This happens, for example, in the  $\{5,4\}$-tiling of $\mathbb{H}^2$. In this case, the presence of the pentagon odd-cycles precludes the possibility of embedding the corner-sharing tetrahedra in two parallel planes as is possible for even $p$. 
These cases, called combinatorial zeolites, correspond to situations without clear geometrical realization, which nonetheless have proved to be interesting from a theoretical point of view \cite{Jordn2016}. Our proposal allows for such layouts to be constructed as planar circuits and their quantum dynamics to be explored.

{
\subsection{Geometrical properties of the circuits}}

{
Even though  we are restricted by construction to planar circuits, 
and consequently to layouts  corresponding to planar graphs, our $q$-leg symmetrical coupling 
can effectively originate quite  generic higher-dimensional Kagom\'e-like structures.
 Let us consider, for the sake of illustration, the case of the octahedral graph $\{3,4\}$, 
\begin{figure}[t]
\includegraphics[width=0.42\linewidth]{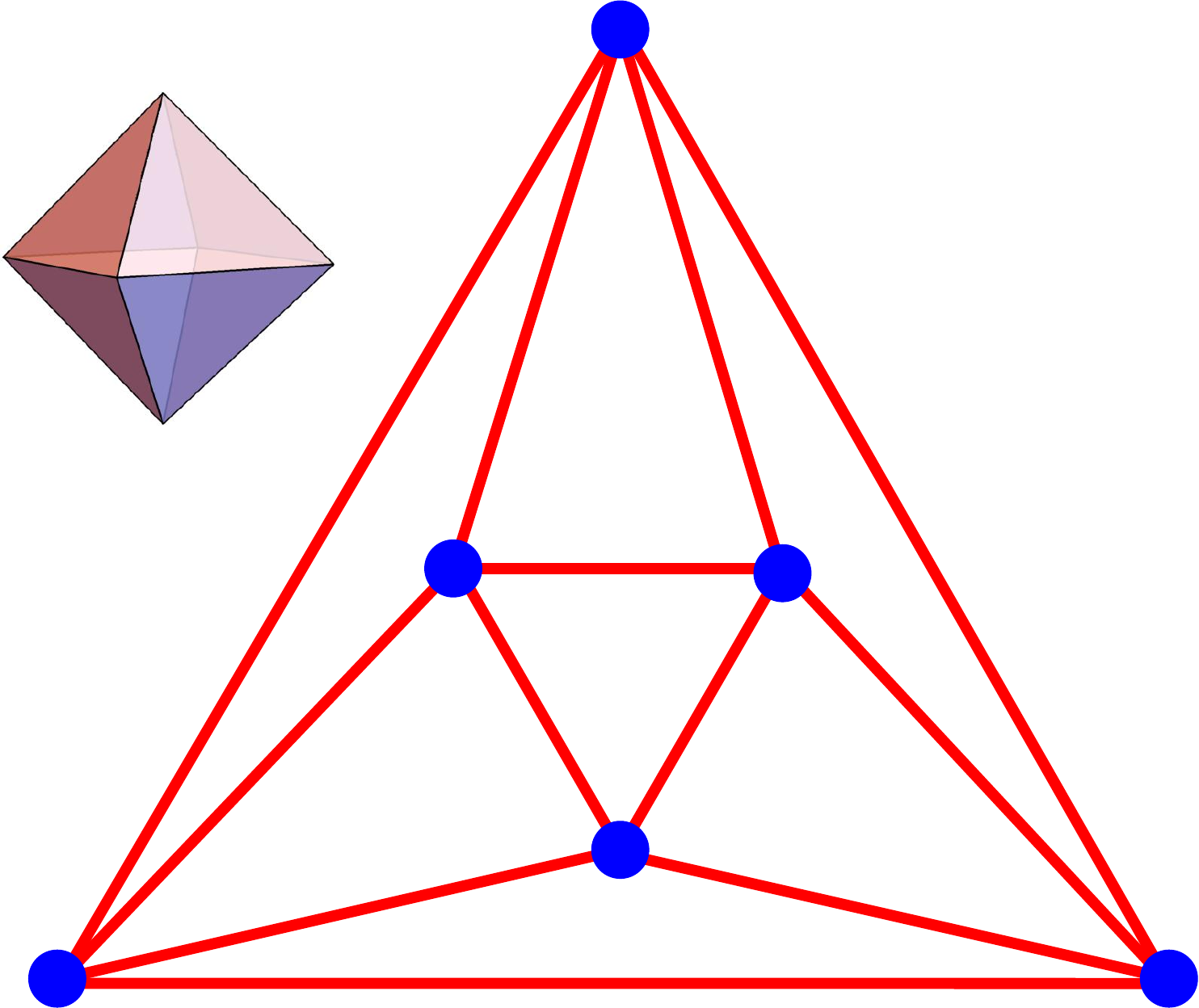}
\includegraphics[width=0.48\linewidth]{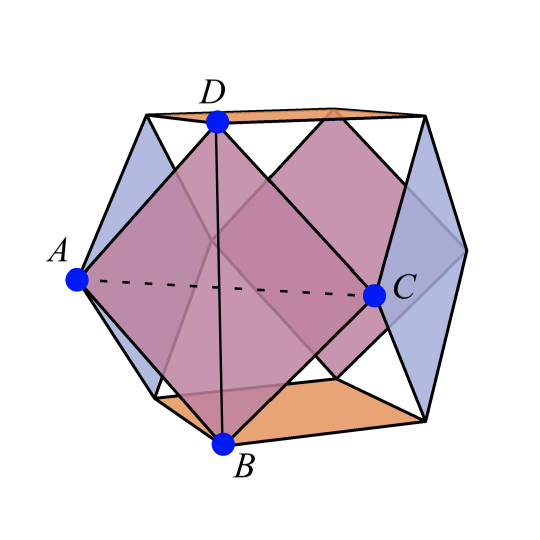}
\caption{\label{octo} {The $\{3,4\}$ tiling of $\mathbb{S}^2$. Left:
the octahedron in  $\mathbb{E}^3$ and its associated planar graph, which can be
implemented as a circuit with symmetrical 4-leg capacitors. Right: 
the cuboctahedron (rectified octahedron) as a
schematic representation of the 
 the octahedron  line graph, which corresponds to 
6 corner-sharing tetrahedra.  
Each square  face of the  cuboctahedron
 is in fact a tetrahedra, but only one is depicted for simplicity. Such structure does not exist in
 $\mathbb{E}^3$, but 
 can be embedded in  $\mathbb{E}^5$, see the text.}
}
\end{figure}  
 which is the simplest 4-regular planar graph
with 6 vertices and 12 edges, corresponding to the 
$\{3,4\}$ tiling of $\mathbb{S}^2$.  
  The associated  Kagom\'e structure in this case  (the  octahedron line graph) 
  is clearly a non planar graph  with 6 pairwise corner-sharing tetrahedra,
see Fig. \ref{octo}.
 However, it is easy to see that it is impossible to assemble such structure in $\mathbb{E}^3$. We necessarily have to go to higher dimensional spaces to get a geometrical representation of the octahedron Kagom\'e structure! The simplest way to embed such structure in a homogeneous space is to consider
 the octahedron not as embedded in $\mathbb{E}^3$, but in $\mathbb{E}^5$, the smallest Euclidean
 space where we can accommodate 6 equidistant points. One possible realization is to locate the vertices
 of the octahedron at the points $(1,0,0,0,0)$, $(0,1,0,0,0)$, $(0,0,1,0,0)$, $(0,0,0,1,0)$, $(0,0,0,0,1)$,
 and $(\varphi,\varphi,\varphi,\varphi,\varphi)$, with 
 $\varphi = \frac{1}{5}\left(1\pm \sqrt{6} \right)$ and connect them accordingly. In this way,
 the resulting line graph will correspond to a configuration with 6 pairwise corner-sharing tetrahedra
 in $\mathbb{E}^5$, which has   effectively arisen   from the planar circuit corresponding to the
 octahedral graph, as depicted in Fig. \ref{octo}. }
 
{
Notice that the inverse problem for these circuits is also well-posed in the sense that, given, for instance, some real or hypothetical \(n\)-zeolite framework, one can in principle determine its equivalent planar layout.
This corresponds to determining the original graph given its line-graph, and such a problem is known to be well-posed in general and it is indeed efficiently implemented in several graph-computing packages \cite{NetworkX}. As an example, it is easy to see that the line-graph of a star graph $S_{\!n+1}$ with $n+1$ leaves is the complete graph $K_{n+1}$, which denotes our $n$-simplex.
Hence, one can determine the circuit equivalent to a certain $n$-zeolite framework substituting the $n$-simplexes with star graphs $S_{\!n+1}$ and connecting the vertices accordingly. This can be illustrated with the
octahedron case of Fig. \ref{octo}. 
It is not difficult to get the original circuits from the corner-sharing $n$-simplex framework   by replacing the $n$-simplexes  by   $S_{\!n+1}$ graphs.}

{
Finally, we stress that since we are restricted to planar layouts, the corresponding graph
circuits will always be embedded in a two-dimensional manifold. In particular, the continuous
limit of \cite{kollar_line-graph_2020} in our case will also give origin to two-dimensional geometries. 
However, due to the design flexibility of our $q$-leg capacitor, we can   explore layouts with spatially varying coordination $q$, which could simulate a non-uniform curvature in a two-dimensional space and, consequently, expand the results of  \cite{kollar_line-graph_2020}   to non-uniform geometries.  }

\subsection{Positive-curvature lattices} 

\begin{figure}[t]
\includegraphics[width=0.49\linewidth]{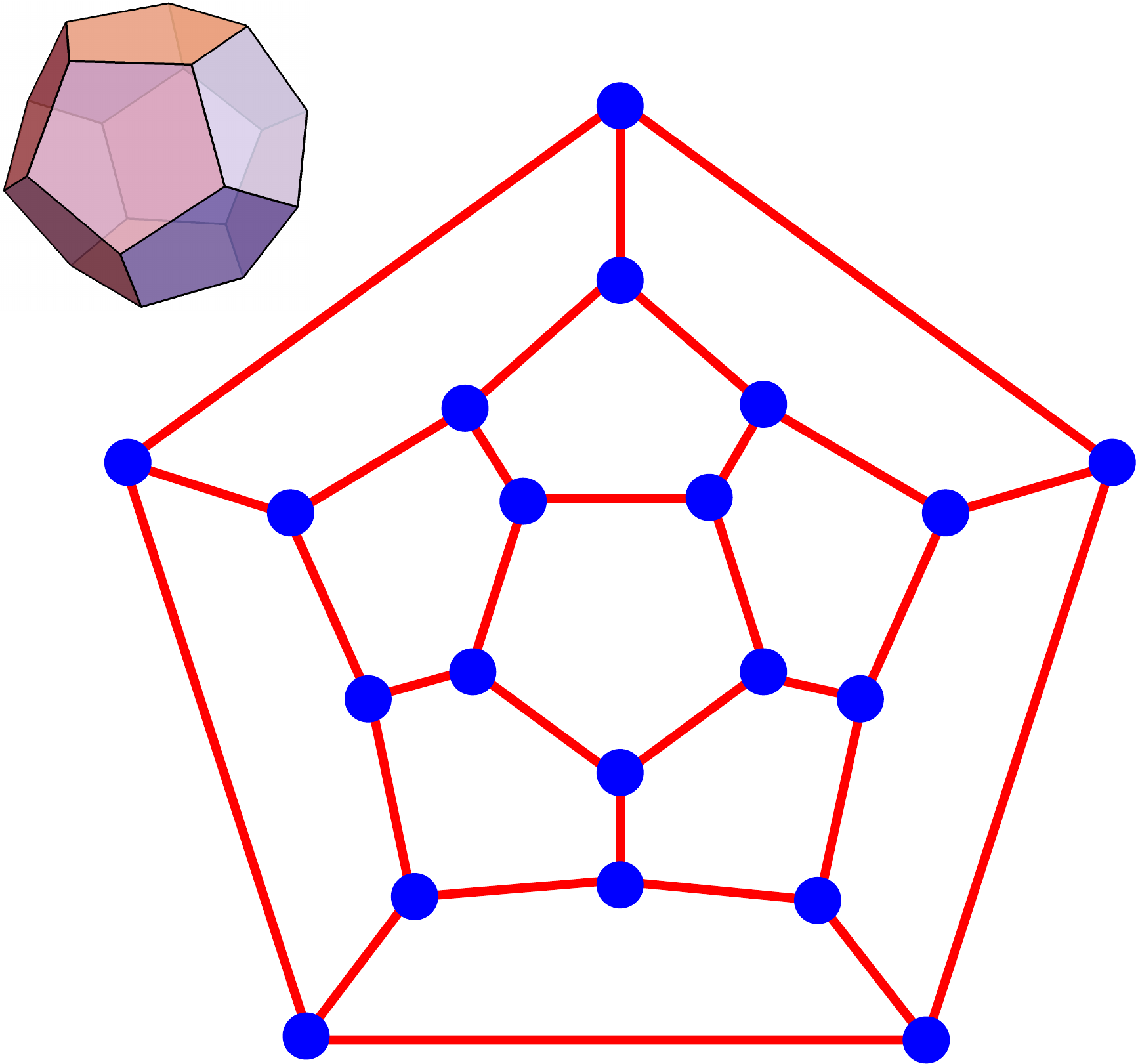}
\includegraphics[width=0.5\linewidth]{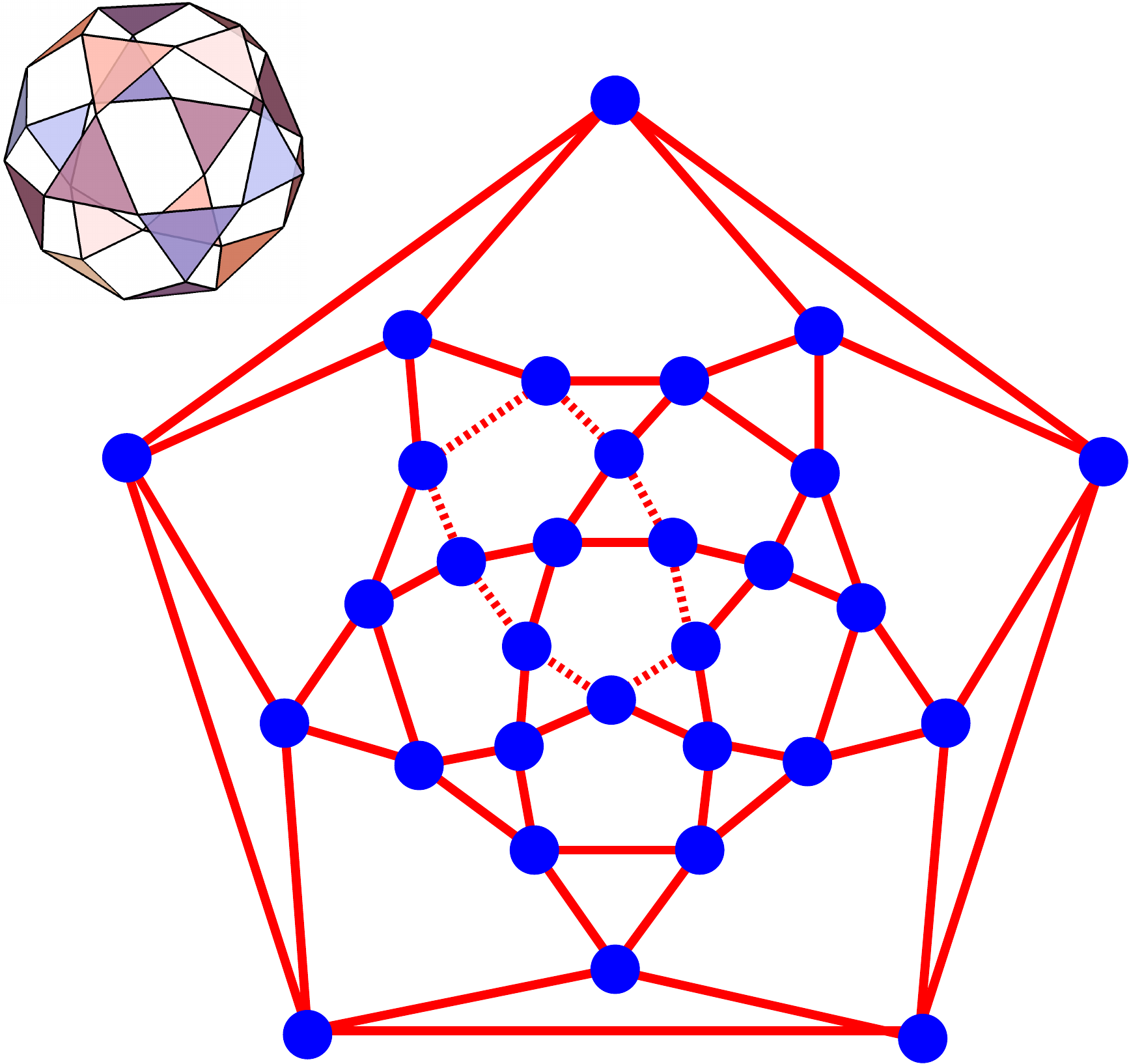}
\caption{\label{fig3} A $\{5,3\}$ tiling of $\mathbb{S}^2$. Left:
the dodecahedron in  $\mathbb{E}^3$ and its planar graph, which can be
implemented as a circuit with symmetrical 3-leg capacitors. Right:  the associated line graph, which is realized as the triangular faces of an icosidodecahedron in  $\mathbb{E}^3$, and its respective 30-vertex graph. The dashed line corresponds to one of the ten even cycles associated with the flat band in the spectra of the icosidodecahedron graph.
}
\end{figure}

It is worth mentioning that even circuits with  $q=3$, as those originally considered in \cite{kollar_hyperbolic_2019}, can also give rise to effectively higher dimensional structures. This is the case, for example, of the fullerenes discussed in \cite{kollar_line-graph_2020}. These correspond to lattices with positive curvature, tilings of the two-sphere $\mathbb{S}^2$, whose embedding requires 3 dimensions. However, both the $C_{60}$ and $C_{84}$ finite tilings of $\mathbb{S}^2$ considered in \cite{kollar_line-graph_2020} involve two different types of faces: pentagons and hexagons. Hence, the associated Kagom\'{e} decoration will necessarily also involve some isosceles triangles besides the equilateral ones associated with the symmetrical capacitor. Although our star-shaped proposal for the capacitor is also able to emulate the isosceles triangles of the associated line graph, one can circumvent this problem by considering the regular dodecahedron circuit shown in Fig.~\ref{fig3},
which can be viewed as the $\{5,3\}$
tiling of the sphere $\mathbb{S}^2$. Since any spherical tiling admits a planar representation, the dodecahedron can be realized as a planar layout circuit, as also shown in 
Fig.~\ref{fig3}. Its line graph is a finite Kagom\'{e} lattice known as an icosidodecahedron (the rectified dodecahedron), a well-known Archimedean solid. This is quite an interesting case to be explored as a circuit due to its amenable size and known analytical spectra.


\section{Some exact results about the spectra}
All the analyses and experiments of \cite{kollar_hyperbolic_2019},  which we propose to extend here, require the knowledge of the excitation spectra of the Hamiltonian of Eq.~\eqref{ham} for both full and half-wave modes. For this, some classical results for finite graphs prove useful. In particular, Lemma 2.1 of   \cite{spectra}   applied to $A_{LG}$ reads
 \begin{equation}
\label{est1}
\chi\left( {A_{LG}}, \lambda\right) = \left(\lambda + 2 \right)^{m-n} \chi\left( Q,\lambda + 2 \right),
\end{equation}
where $\chi\left( M,\lambda  \right)$ denotes the characteristic polynomial for the matrix $M$  in the variable $\lambda$, 
and $Q = D+A$, with $D$, $A$, $n$, and $m$ standing for the degree matrix,  the adjacency matrix, the number of vertices and the number of edges of the \emph{layout} $G$, respectively. The degree of a graph vertex is the number of edges connecting to it (coordination number), and hence the degree matrix here is the diagonal matrix whose entries correspond to the number of resonators connected to each capacitor in the layout circuit.  The matrix $Q$ is known in the graph literature as the signless Laplacian matrix of the graph $G$ (see, e.g., \cite{cvetkovic_signless_2007}).
The same Lemma applied to $A_{LG}^*$ gives
 \begin{equation}
\label{est2}
\chi\left( {A_{LG}^*}, \lambda\right) = \left(\lambda + 2 \right)^{m-n} \chi\left( L,\lambda + 2 \right),
\end{equation}
where $L = D-A$ is the  usual Laplacian matrix of the \emph{layout} $G$. Note how Eqs.~\eqref{est1} and \eqref{est2} relate the spectrum of the line graph $L(G)$ to properties of its layout $G$. Both matrices $Q$ and $L$ are positive semi-definite and, thus, the spectra of both $A_{LG}$ and $A_{LG}^*$ are bounded from below by $-2$. Moreover, there are flat bands with at least $m-n$ eigenvectors with eigenvalue $\lambda_{\rm min}=-2$ for any layout $G$. In fact, for the half-wave modes, the flat band has   $m-n+1$ eigenstates, since $L$ always has a single zero eigenvalue  
due to the fact that the layout is connected \cite{spectra}.  Furthermore, $Q$ also has one vanishing eigenvalue if and only if $G$ is bipartite, in which case we also have $\chi\left( Q,\lambda \right) = \chi\left( L,\lambda  \right)$ \cite{cvetkovic_signless_2007}, so that ${A_{LG}}$ and ${A_{LG}^*}$ have the same spectra. This case corresponds to a layout with a balanced \cite{zaslavsky_signed} signed line graph. Physically, this is a consequence of the fact that the two Hamiltonians corresponding to Eqs.~\eqref{H_s}
and \eqref{H_a} are gauge equivalent in this case.
It is clear from Eqs.~\eqref{est1} and \eqref{est2} that the  spectra of the layout matrices  $Q$ and $L$ suffice to determine the complete spectra of the physical Hamiltonian of Eq.~\eqref{ham} for both full and half-wave modes. All the other eigenvalues belong to the flat band at $\lambda=-2$.

Let us illustrate this with the finite $\{5,3\}$ tiling  of $\mathbb{S}^2$ of
Fig.~\ref{fig3}, whose associated line graph is the icosidodecahedron. The  layout in this case has $L = 3I-A$ and $Q=3I+A$. Moreover, for the dodecahedron \cite{cvetkovic_signless_2007}
 \begin{equation}
 \chi\left( A,\lambda  \right) = \left(\lambda - 3 \right)
 \left(\lambda^2 - 5 \right)^3
 \left(\lambda - 1 \right) ^5
 \lambda^4
 \left(\lambda + 2 \right)^4 ,
 \end{equation}
and, from Eqs.~\eqref{est1} and \eqref{est2},
we have finally the icosidodecahedral graph spectra
\begin{eqnarray}
S\left( {A_{LG}}\right) &=&  \left\{ -2_{10}, \left(1-\sqrt{5}\right)_3, -1_4, 1_4, 2_5, \left(1+\sqrt{5}\right)_3, 4_1 \right\}, \\
S\left( {A_{LG}^*} \right) &=&  \left\{ -2_{11}, \left(1-\sqrt{5}\right)_3, 0_5, 1_4,   \left(1+\sqrt{5}\right)_3, 3_4 \right\},
\end{eqnarray}
where the indices give the respective eigenvalue multiplicities.
One can see that the flat band, which corresponds roughly to 1/3 of the total spectra, effectively comes from the $m-n=10$ term in Eqs.~\eqref{est1} and \eqref{est2}. 
For the full-wave modes, it is quite easy to identify the flat-band eigenvectors: they correspond to an alternating sequence of 1 and $-1$ along even cycles as the one depicted in Fig. \ref{fig3}, and zero elsewhere \cite{kollar_line-graph_2020}.
These cycles are closed paths that go through a unique edge of each visited triangle in the line graph.  
There are 10 linearly independent even cycles of this type in the icosidodecahedron graph, and hence an equal number of $-2$ eigenvalues of $A_{LG}$. Finally, since we are dealing with a regular graph, the largest eigenvalue of  $A_{LG}$ is precisely the line-graph degree, see 
the Appendix \ref{apb}.

\subsection{Flat fraction of the spectra}

For a large $\{p,q\}$-layout, the fraction $f= \frac{m-n}{m}$ ($m\gg 1$) of the spectra corresponding to the flat band is an important property of the circuit. 
We can determine $f$ from the growth properties of the layout graphs (see Appendix \ref{apb} for further details). For large hyperbolic layouts, the flat-band fraction
tends \emph{exponentially} to
\begin{equation}
\label{finf}
 f = \frac{q-2}{\sigma - 1 +q},
\end{equation}
where
\begin{equation}
\label{sigma}
\sigma = \frac{\tau -2 + \sqrt{\tau^2 - 4\tau}}{2},
\end{equation}
with $\tau$ given by Eq.~\eqref{tau}. 
For hyperbolic tilings, $\sigma>1$. Eq.~\eqref{finf} is also valid
for Euclidean tilings (for which $\sigma = 1$) but the convergence is a power law. Spherical tilings are finite and this discussion does not apply.
\begin{figure}
\includegraphics[width=0.45\linewidth]{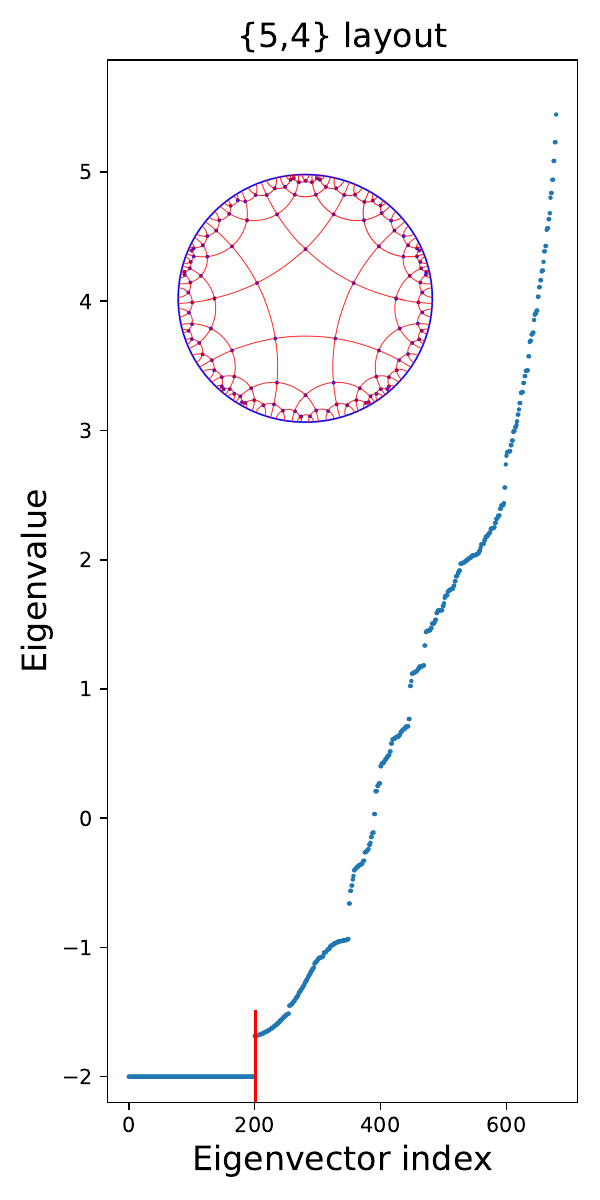}
\includegraphics[width=0.45\linewidth]{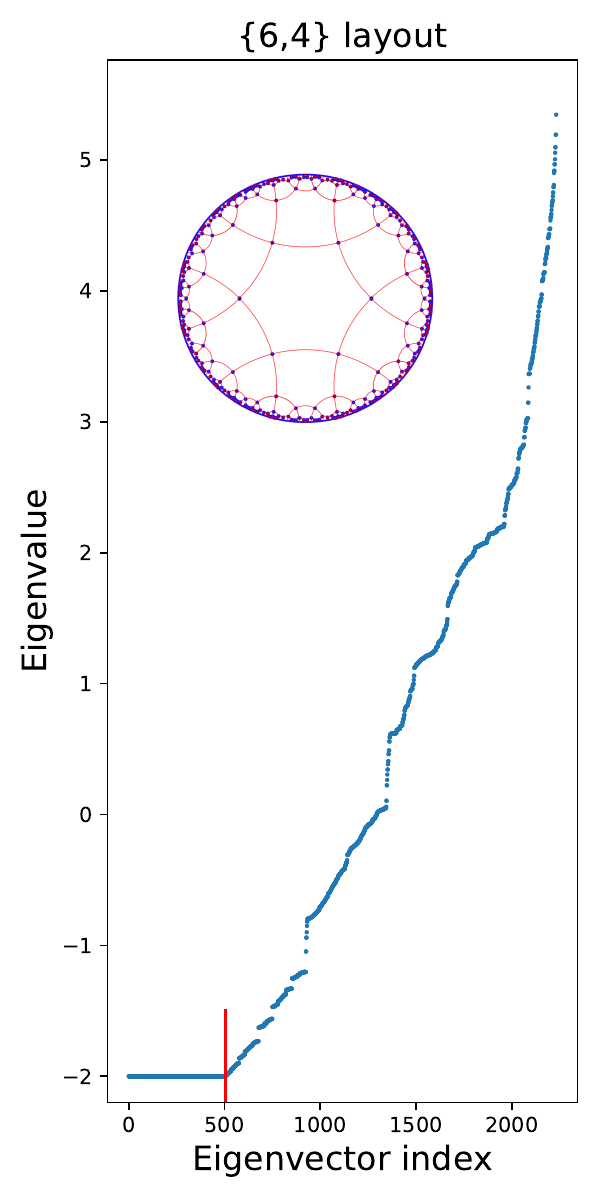}
\caption{\label{fig41} Spectra of the line-graph adjacency matrix $A_{LG}$ for some $\{p,q\}$-layouts, with the red vertical line highlighting the predicted flat-band endpoint.
Left: a layout of 4 concentric rings of the $\{5,4\}$ hyperbolic tilling. The associated line graph has 681 vertices. The predicted flat-band fraction is
$f = 0.297$. Note the gap between the flat band and the rest of the spectra, a property of all layouts with odd $p$. Since $p=5$, this circuit cannot be interpreted as a layer of corner-sharing tetrahedra as in Fig.~\ref{fig1}. Right: a layout of 4 concentric rings of the  $\{6,4\}$ hyperbolic tilling of Fig.~\ref{fig1}. The associated line graph has 2,233 vertices. The predicted flat-band fraction is $f = 0.226$. For even $p$, there is no gap between the flat band and the rest of the spectra.
}
\end{figure}
For the sake of illustration, Fig.~\ref{fig41} 
depicts the spectra for some $\{p,q\}$-layouts. Such spectra are key ingredients in the kinds of experiments performed in \cite{kollar_hyperbolic_2019} and which we propose to extend to $q>3$ configurations.

\section{Conclusions}

In summary, we have shown that, with present-day technology, planar circuit quantum electrodynamics can be explored to simulate some higher-dimensional Euclidean and non-Euclidean structures as, e.g., some $n$-dimensional zeolites, opening the doors to a myriad of new possibilities in metamaterial studies and other related fields. 
We have investigated the spectral properties of line graphs associated with polygon-centered $\{p,q\}$-layouts, with special emphasis on hyperbolic geometries. Using an exact recurrence relation governing the layered growth of these layouts, we derived the flat-band fraction $f_\ell$ and established its exponential convergence to a finite value in hyperbolic tilings, in contrast with the algebraic convergence observed in Euclidean cases. These results provide a precise characterization of the asymptotic spectral weight of the flat band as a function of the local curvature encoded by $p$ and $q$.
We also analyzed the deviation of the average degree $\langle k \rangle$ from the regular value $q$ in finite layouts, showing that the degree deficit is concentrated at the boundary and persists regardless of system size due to the exponential growth of hyperbolic structures. This effect ensures the presence of undercoordinated vertices even in large systems and reinforces the robustness of the flat band.

Finally, motivated by these theoretical insights, we are currently developing superconducting circuits based on compact hyperbolic layouts, notably a dodecahedral configuration. More specifically, we are exploring microwave-guide circuits constructed with sputtered niobium films on silicon substrates. This architecture is designed to probe the spectral features discussed here and to explore experimentally the interplay between geometric frustration, flat-band localization, and nontrivial connectivity. A detailed report on this ongoing work will be provided in a future publication \cite{toppear}. Additionally, another promising avenue for exploration is the study of lattices with spatially varying coordination $q$, which can simulate non-uniform curvature.

The authors acknowledge the financial support of CNPq (Brazil) through grants 302674/2018-7 (AS) and 307041/2017-4 (EM), S\~ao Paulo Research Foundation (FAPESP) through grants 2017/08602-0, 2017/22037-4, 2017/22035-1 (FR).

\appendix

\section{The $q$-leg symmetric capacitor} \label{apa}

We now discuss an efficient implementation of a symmetric planar capacitor with $q$-legs, essential for the cQED application we are proposing. Figure \ref{fig.circuit}  illustrates the schematic geometry of devices featuring \(4\) and \(5\) legs. The \(q\)-leg coupling element, designed as a single central star-shaped section with \(q\)-legs, is positioned at the convergence point of the microwave cavities. Each of these cavities is formed from a segment of a \(Z_0=50\ \Omega\) planar transmission line, coupled at its RF input and output ports through small capacitors \(C_{legs}\). These capacitors set the boundary conditions of the cavity as voltage anti-nodes, facilitating standing-wave resonances with wavelengths \(\lambda = 2L/n\), where \(L\) is the cavity length and \(n\) is an integer. These components can be manufactured using standard micro-fabrication techniques in a single-layer device.

In the weak-coupling limit, where the coupling capacitors \(C_{legs}\), connecting the transmission-line resonators to the \(q\)-leg coupling element, are small compared to the total capacitance of the resonator \(C_{R}\), the \(q\)-leg elements can be adiabatically eliminated \cite{koch_time-reversal-symmetry_2010, nunnenkamp_synthetic_2011}, allowing the system to be effectively described by a tight-binding Hamiltonian, as shown in Eq. (\ref{ham}). The photon hopping amplitude between two resonators is then \cite{koch_time-reversal-symmetry_2010,nunnenkamp_synthetic_2011}
\begin{equation}
t_{ij} \propto C_{legs}\phi_{ij}
\end{equation}
where \(\Phi_{ij}\) is the voltage mode function of the pair \((i,j)\).

{To ensure homogeneous photon hopping amplitudes, the capacitance between any two resonators \( (i, j) \) within the network must be the same. To demonstrate the viability of constructing such devices, we simulated the capacitance between the cavities depicted in Figure \ref{fig.circuit} using the Ansys Q3D Extractor software. This process involved utilizing the CAD file of our circuit to solve Maxwell's equations, thereby determining the field and charge distributions. Employing standard parameters used in cQED devices, we conducted electrostatic simulations of the $q$-leg geometry based on the layout illustrated in Figure \ref{fig.circuit}. These simulations were performed for devices on a 500 $\mu$m silicon substrate with a relative permittivity of 11.45, and the superconducting thin film was modeled as a 100 nm thick perfect electric conductor. The resulting capacitance between any two legs was determined to be \(0.37399\pm0.00001\) fF and \(0.27110\pm0.00003\) fF for the \(4\)-leg and \(5\)-leg configurations, respectively. These results suggest that the proposed geometry for the \(q\)-leg capacitor can achieve uniform photon hopping within the circuit.}

\section{Spectra and growth properties of layouts}
\label{apb}

\begin{figure}[t]
\includegraphics[width=0.6\linewidth]{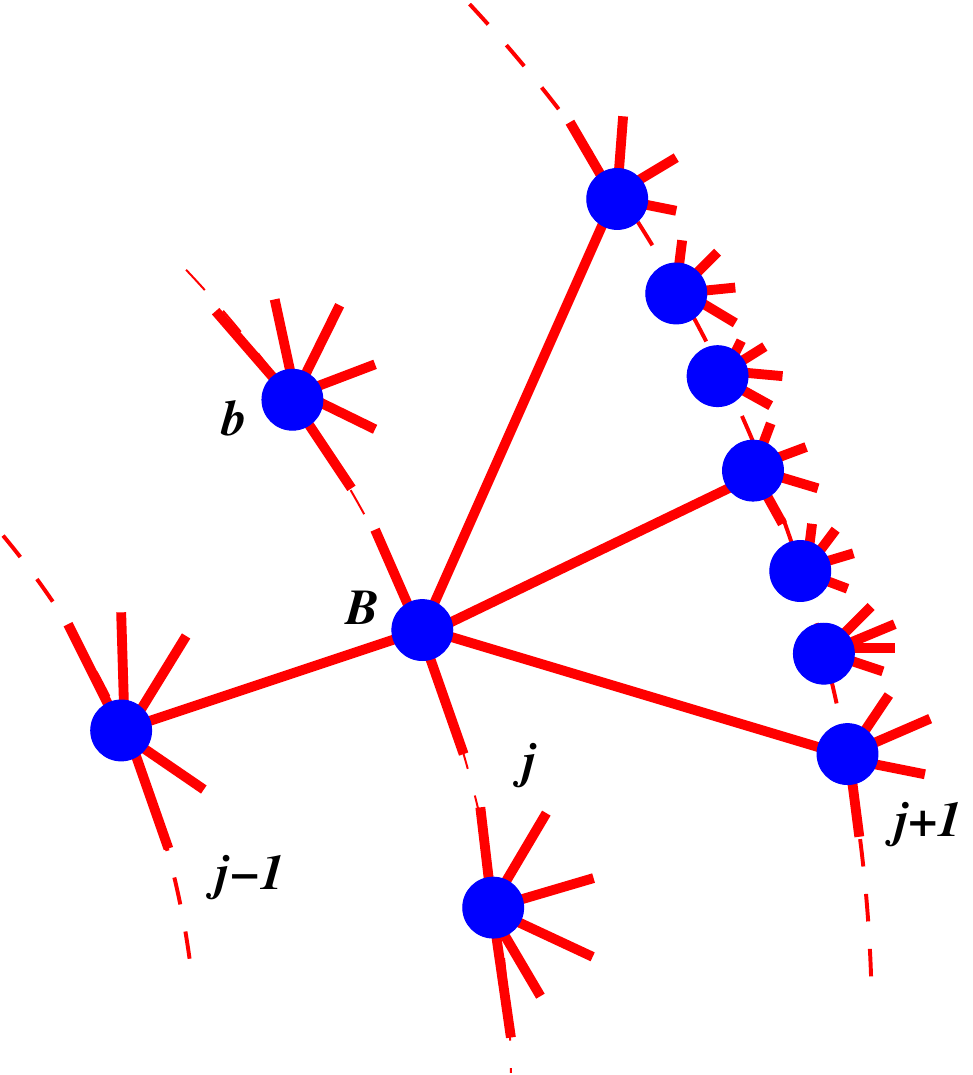}
\caption{\label{fig4} Three concentric rings of a polygon-centered $\{p,q\}$-layout ($p> 3$) with their two types of vertices: the $B$-type,  connecting the $j^{\rm\, th}$ ring to the previous $(j-1)^{\rm\, th}$ one, and the $b$-type, that do not connect to the previous ring. There are $q$ edges meeting at every vertex. }
 \end{figure}

\begin{figure}[ht]
\includegraphics[width=0.95\linewidth]{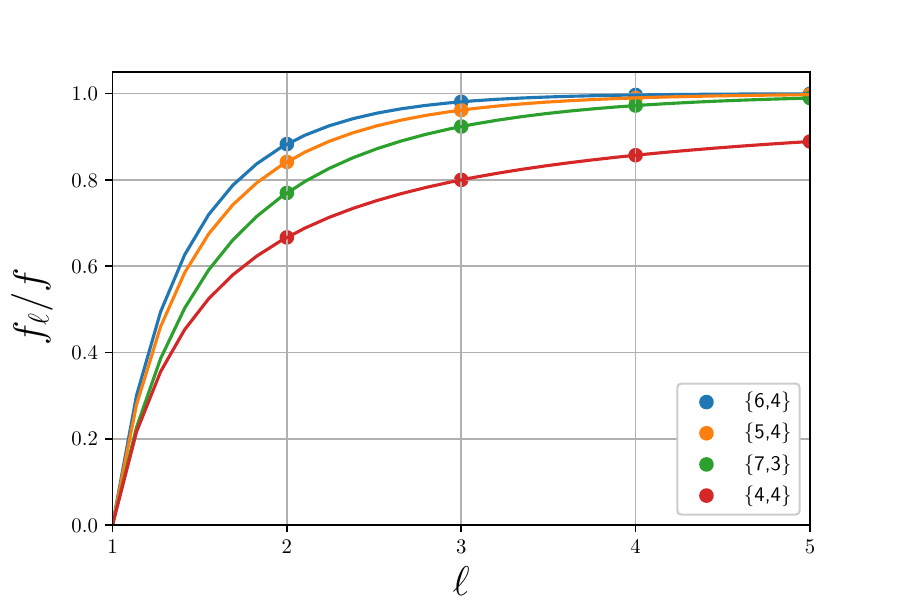} 
\caption{\label{grow}  Convergence of $f_\ell/f$ as a function of the number of rings $\ell$ of the layout [see Eqs. (\ref{finf})  and (\ref{fell})] for different tilings. The convergence for
hyperbolic tilings is exponential, in contrast to the power-law convergence for the Euclidean case ($\{4,4\}$).}
 \end{figure}

The fraction $ f= \frac{m-n}{m}$ corresponding to the proportion of zero eigenvalues in the spectrum of the adjacency matrix of the line graph, which corresponds to the flat band,   is a key feature in the spectral analysis of these circuits.
Recalling that the average degree  $\langle k \rangle $ of a graph with $m$ edges and $n$ vertex is given by
\begin{equation}
\langle k \rangle = \frac{2m}{n}    ,
\end{equation}
we have
\begin{equation} 
\label{frac}
f= 1 - \frac{2}{\langle k \rangle}.
\end{equation}
We can  obtain the fraction $f$ for finite $\{p,q\}$-hyperbolic layouts from the growth properties of these graphs.
The problem of the growth of vertex-centered hyperbolic tilings was considered in \cite{moran_growth_1997}. One can  easily adapt that approach to our problem of growing polygon-centered layouts by the accretion of concentric layers of tilings.
Let us assume we have a layout composed of $\ell$ concentric rings of vertices, ordered outwards, of a $\{p,q\}$-tiling, with $p>3$. It will become clear that the case of a triangular tiling ($p=3$) is intrinsically different and will not be treated here since it does not seem to be interesting for our purposes.
Each ring $j$ contains two types of vertices: type-$B$ vertices, which connect to the previous $(j-1)^{\rm th}$ ring, and type-$b$ vertices, which do not. These are illustrated in Fig.~\ref{fig4}.
Let $b_j$ be the number of vertices on the  $j^{\rm\, th}$ ring that are not connected to the $(j-1)^{\rm\, th}$ one, and $B_j$ the number of remaining vertices  which are connected to previous ring. For example, for the $\{6,4\}$ tiling of Fig. \ref{fig1}, one has $b_1=6$, $B_1=0$, $b_2 =30$,  $B_2= 12$, and so on. Each edge  emanating from the $j^{\rm\, th}$ ring will necessarily reach a $B$-type vertex in the $(j+1)^{\rm\, th}$ ring and, thus, we have
\begin{equation}
\label{recurrence1}
B_{j+1} = (q-2)b_j + (q-3)B_j.
\end{equation}
The recurrence for the $b$-vertices is a little more intricate. From Fig.~\ref{fig4}, we see that for each $B$-vertex, there are $q-2$ polygons between the $j^{\rm\, th}$   and the $(j+1)^{\rm\, th}$ rings. For the $b$-vertices, there are $q-1$ of such polygons. Each one of these polygons, which we assume to be ordered anticlockwise,  will lead to $p-3$ $b$-vertices in the $(j+1)^{\rm\, th}$ ring. To compute $b_{j+1}$, we run circularly over all these polygons  between the $j^{\rm\, th}$ and the
$(j+1)^{\rm\, th}$ rings. In order to avoid double counting, we neglect the last polygon of each vertex, since it coincides with the first one of the next vertex. We must also neglect one vertex in the sum of each vertex in the  $j^{\rm\, th}$, since the first polygon, in contrast to the other ones (with the exception of the last), has one of its edges on the  $j^{\rm\, th}$  ring.
Finally, we have the following recurrence system, valid for $p>3$,
\begin{equation}
\label{recurrence}
\left(
\begin{matrix}
b_{j+1} \\
B_{j+1}
\end{matrix}
 \right) =
 \left(
\begin{matrix}
(q-2)(p-3) -1 & (q-3)(p-3)-1\\
q-2 & q-3
\end{matrix}
 \right)
 \left(
\begin{matrix}
b_{j} \\
B_{j}
\end{matrix}
 \right)  .
\end{equation}
For any polygon-centered $\{p,q\}$-layout, the initial condition for Eq.~(\ref{recurrence}) is $b_1=p$ and $B_1=0$. We can determine the number of edges $m_\ell$ and  vertices $n_\ell$ of a $\{p,q\}$-layout consisting of $\ell$ concentric rings from the function $B_\ell$ alone. Following \cite{moran_growth_1997}, let $t_\ell$ be the number of polygons in the layout.
Then,
\begin{equation}
t_\ell = 1 +  \sum_{j=1}^\ell B_j.
\end{equation}
The number of vertices in the same layout will be given by
\begin{equation}
n_\ell = \sum_{j=1}^\ell\left(b_j+B_j \right) = \frac{1}{q-2}\sum_{j=1}^\ell \left( B_{j+1} + B_j \right) =
\frac{B_{\ell+1} + 2( t_\ell -1)}{q-2},
\end{equation}
where Eq.~\eqref{recurrence1} was used.
The number of edges $m_\ell$ can be determined from Euler's formula for planar graphs
\begin{equation}
n_\ell - m_\ell + t_\ell = 1,
\end{equation}
from which we finally have the fraction
\begin{equation}
\label{fell}
f_\ell = \frac{m_\ell - n_\ell}{m_\ell} = \frac{q-2 }{C_{\ell} + q },
\end{equation}
where
\begin{equation}
C_\ell = \frac{B_{\ell+1}}{t_\ell - 1}.
\end{equation}
The   fraction of Eq.~\eqref{fell} for large layouts is determined by $\lim_{\ell \to\infty }C_\ell$. In order
to evaluate this limit, let us consider the equation for $B_\ell$ obtained from the recurrence system of Eq.~\eqref{recurrence}
\begin{equation}
\label{A7}
B_{\ell+1} = (\tau -2)B_\ell - B_{\ell - 1},
\end{equation}
with $\tau$ given by Eq. (\ref{tau}),
whose solution for our case is
\begin{equation}
\label{A8}
B_\ell =  \frac{p(q-2)}{\sigma^2 -1}\left(\sigma^\ell - \sigma^{2-\ell} \right),
\end{equation}
with $\sigma$ given by Eq. (\ref{sigma}). Note that this solution is valid only for hyperbolic tilings. For Euclidean ones $\sigma = 1$ and the solution is $B_\ell = p(q-2)(\ell-1)$.  From Eq.~\eqref{A8},
\begin{equation}
t_\ell  =  1+\frac{p(q-2)}{\sigma^2 -1}  \frac{\sigma^{\ell+1} -\sigma^2- \sigma + \sigma^{2-\ell}}{\sigma - 1}   ,
\end{equation}
yielding
\begin{equation}
C_\ell =  \frac{\left( \sigma -1 \right)\left(1 - \sigma^{ -2\ell}\right)}{1 + \sigma^{1-2\ell} -  (\sigma + 1)\sigma^{-\ell}},
\end{equation}
and finally
\begin{equation}
\lim_{\ell\to\infty} C_\ell = \sigma - 1,
\end{equation}
from which Eq. (\ref{finf}) follows immediately. 
For Euclidean tilings, we have instead
\begin{equation}
C_\ell = \frac{2}{\ell - 1},     
\end{equation}
which is also compatible with Eq. (\ref{finf}), albeit with a slower power-law convergence. 
Fig. \ref{grow}
illustrates the convergence of $f_\ell$ as a function of the number
of rings $\ell$ of the layout for different tilings.
 
It is worth  mentioning that from Eqs. (\ref{finf})  and \eqref{frac}, the average degree of a large $\{p,q\}$-layout is
\begin{equation}
\langle k \rangle = 2 \left(\frac{\sigma - 1 + q}{\sigma + 1}\right).
\end{equation}
This shows that, although hyperbolic tilings are $q$-regular, we always have  $\langle k \rangle < q$ for any finite hyperbolic layout, no matter how large it is. This is hardly surprising since all vertices with degree deficiency $(k < q)$  are located in the outermost ring of the layout and hyperbolic tilings grow exponentially. In contrast,  Euclidean tilings grow linearly and have  $\langle k \rangle = q$.

Besides the flat band, we can also estimate the largest eigenvalues of $A_{LG}$  and $A_{LG}^*$ from some classical results for the spectra of the matrices   $Q$ and $L$.   For instance, if $\mu$ stands for the largest eigenvalue of $Q$, one has  \cite{cvetkovic_signless_2007}
$
2 k_{\rm min} \le \mu \le 2 k_{\rm max},
$
where $k_{\rm min}$ and $k_{\rm max}$ stand for, respectively, the minimal and maximal degree of the layout, with the equality holding if and only if $G$ is regular. For our case, 
$k_{\rm min} = 2$ in the outermost ring and $k_{\rm max} = q$, implying
\begin{equation}
2 \le \max [S\left({A_{LG}}\right)] \le 2(q-1).
\end{equation}
There are many similar bounds for the largest eigenvalue of the Laplacian matrix, and they can be used to estimate the largest eigenvalues of $A_{LG}^*$ analogously.  For instance, from the elementary bound \cite{spectra} $ k_{\rm max} \le \nu \le 2 k_{\rm max}$ for 
the the largest eigenvalue $\nu$ of $L$, we have
\begin{equation}
q-2 \le \max [S\left({A_{LG}^*}\right)]\le 2(q-1).
\end{equation}
These bounds can be checked against Fig. \ref{fig41}.

\bibliographystyle{apsrev4-2}


%
 
\end{document}